\documentclass[twocolumn]{aastex61}
\usepackage{multirow}
\usepackage{natbib}
\usepackage{array}
\usepackage{float}
\usepackage{xcolor}
\usepackage{graphicx}
\usepackage{hyperref}
\usepackage{epstopdf}
\newcolumntype{L}[1]{>{\raggedright\let\newline\\\arraybackslash\hspace{0pt}}m{#1}}
\newcolumntype{C}[1]{>{\centering\let\newline\\\arraybackslash\hspace{0pt}}m{#1}}
\newcolumntype{R}[1]{>{\raggedleft\let\newline\\\arraybackslash\hspace{0pt}}m{#1}}
\definecolor{myred}{rgb}{1.0,0.40,0.40}

\newcommand{\kms}{km\,s$^{-1}$}

\received{April 26, 2019}
\accepted{June 3, 2019}
\submitjournal{ApJ}

\shorttitle{Spectroscopy and DEM diagnostics  of a coronal dimming }
\shortauthors{Veronig et al.}

\begin{document}

\title{Spectroscopy and Differential Emission Measure diagnostics  of a coronal dimming associated with a fast halo CME}
\correspondingauthor{Astrid M. Veronig}
\email{astrid.veronig@uni-graz.at}

\author{Astrid M. Veronig}
\affil{Institute of Physics, 
University of Graz,
8010 Graz, Austria}
\affil{Kanzelh\"{o}he Observatory for Solar and Environmental Research,
University of Graz,
9521 Treffen, Austria}

\author{Peter G\"{o}m\"{o}ry}
\affil{Astronomical Institute of the Slovak Academy of Sciences,  
05960 Tatransk\'a Lomnica, 
Slovakia}

\author{Karin Dissauer}
\affil{Institute of Physics, University of Graz, 8010 Graz, Austria}

\author{Manuela Temmer}
\affil{Institute of Physics, University of Graz, 8010 Graz, Austria}

\author{Kamalam Vanninathan}
\affil{Institute of Physics, University of Graz, 8010 Graz, Austria}

\begin{abstract}
We study the coronal dimming caused by the fast halo CME (deprojected speed  $v =1250$ \kms) associated with the C3.7 two-ribbon flare on 2012 September 27, using Hinode/EIS spectroscopy and SDO/AIA Differential Emission Measure (DEM) analysis. The event reveals bipolar core dimmings encompassed by hook-shaped flare ribbons located at the ends of the flare-related polarity inversion line, and marking the footpoints of the erupting filament. In coronal emission lines of $\log T \, [{\rm K}] = 5.8-6.3$, distinct double component spectra indicative of the superposition of a stationary and a fast up-flowing plasma component with velocities up to 130 \kms\ are observed at regions, which were mapped by the scanning EIS slit close in time of their impulsive dimming onset. The outflowing plasma component is found to be of the same order and even dominant over the stationary one, with electron densities in the upflowing component of $2\times 10^{9}$\,cm$^{-3}$ at $\log T \, [{\rm K}]  = 6.2$. The density evolution in core dimming regions derived from SDO/AIA DEM analysis reveals impulsive reductions by 40--50\% within $\lesssim$10 min, and remains at these reduced levels for hours. The mass loss rate derived from the EIS spectroscopy in the dimming regions is of the same order than the  mass increase rate observed in the associated white light CME ($1 \times 10^{12} {\rm \; g \; s}^{-1}$), indicative that the CME mass increase in the coronagraphic field-of-view results from plasma flows from below and not from material piled-up ahead of the outward moving and expanding CME front. 
\end{abstract}

\keywords{sun: atmosphere --- sun: corona  --- sun: coronal mass ejections (CMEs) --- sun: flares --- methods: observational --- techniques: spectroscopic}

\section{Introduction}
\label{sec:intro}

Coronal dimmings are regions that undergo an abrupt reduction in the solar Extreme-Ultraviolet (EUV) and/or soft X-ray emission, co-temporal with the launch of a coronal mass ejection (CME). ``Abrupt depletions'' of localized regions in the inner corona were first discussed by \cite{Hansen1974} based on Mauna Loa white-light coronagraph data in association with an H$\alpha$ prominence eruption, and were already presumed to be due to the expulsion of coronal material. 

Observations of coronal dimmings in soft X-rays with the Yohkoh SXT  \citep{Hudson1996, Sterling1997} and in the EUV with the Extreme-ultraviolet Imaging Telescope (EIT) onboard the Solar and Heliospheric Observatory \citep[SOHO;][]{Thompson1998,Zarro1999} reinforced the interpretation that the reduced emission in coronal dimmings is a result of a density depletion caused by expansion and evacuation of coronal plasma due to the erupting CME \citep[e.g.,][]{Hudson1996,Harrison2000,Zhukov2004}. Coronal dimmings represent the most distinct phenomena associated with CMEs low in the corona, and contain important information on their initiation and early evolution, before the CME is observed in coronagraphs. In a recent series of papers, coronal dimmings and the associated CMEs were extensively studied in co-temporal observations from  quasi-quadrature view using the Atmospheric Imaging Assembly (AIA) onboard the Solar Dynamics Observatory (SDO) and the Extreme-Utraviolet Imager (EUVI) and COR instruments onboard the Solar-Terrestrial Relations Observatory (STEREO). These studies revealed that the maximum extent, the magnetic flux and the amplitude of the intensity reduction of the dimming region show distinct correlations with the CME mass, whereas the dynamics of the dimming evolution (area growth rate, intensity change rate) is correlated with the CME speed \cite[][]{Dissauer2018a,Dissauer2018b,Dissauer2019}.

Coronal dimmings are often differentiated into two categories: core and secondary dimmings \citep{Mandrini2007,Dissauer2018a}. Core (or twin) dimmings are localized areas of strongly reduced emission, located close to the eruption site and rooted in regions of opposite magnetic polarity regions. They are interpreted as marking the footpoints of the erupting flux rope \citep{Hudson1996,Zarro1999,Webb2000} and may reveal de-sharing motions similar to the flare ribbon evolution \citep{Miklenic2011}, while the more shallow secondary dimmings are understood to be due to the overlying field that is stretched and partly reconnecting \citep{Mandrini2007,Attrill2009,Thompson2000}. 
Using Differential Emission Measure (DEM) analysis  on the multiband AIA EUV imagery, \cite{Vanninathan2018} found significant differences in the plasma properties of the core and the secondary dimming regions, in terms of their depletion depth, rate and refill time. In the core dimmings, the density dropped impulsively within $<$30 minutes by up to 50--70\%, and thereafter stayed at such low levels for more than 10~hours, while the secondary dimmings evolve more gradually with less strong depletions, and start to refill already 1--2 hours after the start of the event. 

Observations of plasma outflows in dimming regions give further evidence that these regions are a result of mass loss and density depletion. However, due to the transient and localized nature of coronal dimmings and the limited spatio-temporal coverage of spectrometers, there exists only a few spectroscopic studies of coronal dimmings. \cite{Harrison2000} made the first spectroscopic observation of dimmings using the Coronal Diagnostic Spectrometer (CDS) onboard SOHO, and showed that the mass loss in dimmings could account for up to 70\% of the CME mass. \cite{Harra2001} reported significant blueshifts in coronal and transition region lines in CME-associated dimmings, indicative of mass outflows.
\cite{Harra2007} presented the first spectroscopic studies of a coronal dimming associated with a halo CME using the Extreme-ultraviolet Imaging Spectrometer (EIS) onboard Hinode. They find that the outflows in the dimming regions are highly structured and concentrated in extended loops, with the strongest outflows located at the footpoints of these  loops. As shown in \cite{Jin2009}, the outflow speeds correlate with the underlying photospheric field strength as well as the magnitude of the dimming. 

The plasma outflows in coronal dimming regions are observed  at different heights, from the low transition region to the corona
\citep[e.g.,][]{Jin2009}. As discussed in \cite{Tian2012}, spectral lines in coronal dimming regions reveal significant blueshifts and enhanced line widths.  The outflow speeds  derived from single Gaussian line fits lie mostly in the range of 10--40 \kms.
Studying several coronal dimmings, which were well covered in spectroscopic observations by Hinode/EIS, \cite{Tian2012} show that the coronal emission line profiles in coronal dimmings actually indicate that these are caused by the superposition of a strong background emission component and a relatively weak (of the order of 10\%) upflow component with high speeds of $\approx$100 \kms.

In a few events, temperature dependent outflows have been observed. \cite{Imada2007} report a case of temperature-dependence in the outflow speeds on the border of a dimming region. The  speed of the flow increased from about 10 \kms at $\log T \, [{\rm K}]  = 4.9$ up to 150 \kms  at $\log T \, [{\rm K}]  = 6.3$. \cite{Tian2012}
found that outflows generally do not show any temperature dependence in the majority of the dimming regions and that the temperature-dependent outflows are found only at small regions immediately outside the (deepest) dimming regions.  

Finally, we note that the outflows observed from coronal dimmings may be also a relevant parameter in understanding the evolution of CME mass. It has been reported that the CME mass increases from the low corona to 1 AU \cite[e.g.,][]{Webb1996,Vourlidas2000,Tappin2006,DeForest2013}. Studies of the CME in white-light coronagraph data in the low to mid corona suggest that this increase is predominantly coming from the regions {\em behind} the CME and not due to pile-up (``snow-plough") of material ahead of the CME front \citep{Vourlidas2010,Bein2013,Feng2015,Howard2018}. A viable candidate to explain such feeding of mass to the evolving CME from below would be mass flows from the coronal dimming regions \cite[e.g.,][]{Temmer2017}.

In this paper, we study Hinode/EIS spectroscopy of a core dimming region that resulted from a fast halo CME, in combination with plasma diagnostics of the dimming region and its dynamical evolution using DEM analysis derived from the SDO/AIA EUV filtergrams. As we will demonstrate, localized regions that are ``activated'' as dimming pixels close in time when mapped by the EIS slit, reveal clear double Gaussian line profiles, where the amount of plasma in the upflowing component can be even larger than that in the stationary background plasma. The estimates of the density and speed of the upflowing plasma will be set in comparison with the increase of mass of the associated CME derived from the STEREO coronagraphs  up to a distance of 15 solar radii.

\section{Data and Observations}
\label{sec:obs}

The coronal dimming studied in this paper is associated with an eruptive C3.7 class flare which occurred  on 27 September 2012 at (N09$^\circ$,W33$^\circ$) in NOAA Active Region 11577. The start of the GOES soft X-ray flare was at 23:36\,UT with the maximum reached at 23:57\,UT. Despite being only of GOES class C3.7, the event was an extended two-ribbon flare associated with a fast halo CME with a mean projected plane-of-sky speed of about 950 \kms, as listed in the LASCO CME catalogue 
\cite[\url{https://cdaw.gsfc.nasa.gov/CME_list/};][]{Yashiro2004}. Notably, this CME was associated with a magnetic cloud measured in-situ at L1 on 2012 October 1 (start time around 00:00 UT)\footnote{See Richardson and Cane ICME list, \url{http://www.srl.caltech.edu/ACE/ASC/DATA/level3/icmetable2.htm}}, and produced a strong geomagnetic storm with a Dst of $-122$ and Kp of $7-$.

The full-disk multiband EUV images recorded by the Atmospheric Imaging Assembly \citep[AIA;][]{Lemen2012} on-board the Solar Dynamics Observatory \citep[SDO;][]{Pesnell2012} captured the evolution of this event with high spatial and temporal resolution. The Extreme-Ultraviolet Imaging Spectrometer \citep[EIS;][]{Culhane2007} on-board Hinode \citep{Kosugi2007} was pointed at NOAA 11577, and captured part of the Western flare ribbons as well as the associated Western core dimminig region. Thus, this event provides us with an excellent opportunity to study the plasma flows and densities in coronal dimming regions spectroscopically (with EIS) as well as via Differential Emission Measure diagnostics (with AIA).

The AIA instrument onboard SDO observes the solar corona in six EUV wavelength bands, centered at 94, 131, 171, 193, 211, and 335 {\AA}, thus providing plasma diagnostics over a temparature range of $\log T \, [{\rm K}]  =5.7-7.0$ \citep{Lemen2012}. In addition, AIA observes the lower transition region and upper chromosphere in the He\,{\sc ii} line centered at $\log T \, [{\rm K}] = 4.7$. The observing cadence of the AIA EUV channels  is 12 s and the pixel scale of the CCD is 0.6\arcsec. 

The EIS observations  of the event consist of one raster taken between 23:10:27\,UT and 00:20:36\,UT, covering an area of $302.5\arcsec\times384.0\arcsec$. 
%(cf.\ the box in the second AIA 211{\AA} image in Figure \ref{fig:overview}). 
The scan was in 101 steps (from right to left) using a slit width of 2\arcsec, slit height of 384\arcsec\ and a pixel size of 1\arcsec\ in the $y$-direction. The exposure time for each slit position was 40\,s. In total, 25 spectral windows were registered, and the observed spectral lines cover a wide range of temperature  of 
$\log T \, [{\rm K}]  =4.7-7.2$. In this study we focus on the spectral lines where the dimming region is clearly visible and can be compared with AIA images. For this purpose, we selected the following EIS spectral lines: Fe\,{\sc xiii} at 196.55\,\AA\ and 202.04\,\AA\ ($\log T \, [{\rm K}]  = 6.2$), Fe\,{\sc xv} at 284.16\,\AA\ ($\log T \, [{\rm K}]  = 6.3$), and Si\,{\sc vii} at 275.37\,\AA\ ($\log T \, [{\rm K}]  = 5.8$). The Fe\,{\sc xiii} line ratios are used for density measurements. Lines with signature of blends were not used in this study, which includes the strongest line Fe\,{\sc xii} at 195.12\,\AA.
We note that the Fe\,{\sc xv} line at 284.16\,\AA\ is actually blended with the Al\,{\sc ix} 284.03 line. However, as shown in \cite{Young2007}, the Al\,{\sc ix} line at 284.03 A becomes only apparent in quiet Sun conditions where the Fe\,{\sc xv} emission is very weak. In our observations of AR 11577, the signal is strong and dominated by the Fe\,{\sc xv} emission.

The compensation of all known photometric effects in the EIS spectral data and their calibration to physical units was performed using the eis\_prep.pro routine, which is part of the SolarSoftWare. The wavelength drift was corrected by the method described in \cite{Kamio2010}. As a first approximation, the spectral profiles of the selected lines were fitted by a single Gaussian function with a linear background in order to determine the spectral characteristics  of the observed profiles, i.e., spectral amplitude, integrated intensities, background intensities, Doppler shifts, and spectral widths. 
The EIS data provide no absolute wavelength calibration, and thus the zero reference of the Doppler shifts was calculated as the average value of the Doppler shifts from quiet-Sun regions which were not directly affected by the flare or dimming. 

The EIS data recorded at different wavelengths are affected by a spatial offset in $y$-direction. This offset was compensated for the spectral lines under study. Thus the spatial pixels of the different EIS spectral windows correspond to the same area on the Sun. Finally, the acquired EIS raster was co-aligned with the AIA filtergrams. For this purpose, we used an AIA image recorded in the 211\,\AA\ filter to co-align with the EIS Fe\,{\sc xiii} 202.04\,\AA\ raster, as they both are probing solar plasmas at similar temperatures. We have chosen the AIA image recorded on 2012 September 28 at 00:06:11 UT, as this corresponds to the time when the EIS slit was scanning the inner part of the triangular-shaped Western dimming region, in which we are mostly interested.
The position of the EIS raster relative to the AIA image was calculated by applying spatial cross-correlation on the two images, and we obtained a shift between the nominal EIS and AIA coordinate systems of $x=11.5$\arcsec\ and $y=9.0$\arcsec. These corrections were applied to the EIS data, and the coordinate system of AIA was used as reference.  

Apart from spectroscopy, Differential Emission Measure (DEM) analysis on filter images provide an alternative method to retrieve information on the plasma parameters. The added value of the AIA DEM analysis lies in the possibility of studying the dynamic evolution of the plasma parameters in the dimming regions, whereas EIS provides us with a snapshot of the full spectroscopic information. In the present case, we have only one EIS scan avaliable during the event, and no information on the pre-event coronal state.
We have used the regularized inversion method of \cite{Hannah2012,Hannah2013} to derive the DEM from the six AIA EUV filters for every pixel of the image. The AIA images and thus the reconstructed DEM maps have a pixel resolution of 0.6\arcsec. 
%For the DEM analysis, we first binned the multiband AIA data to $4\times 4$ pixels using flux conservation to obtain a better signal-to-boise ratio, which is in particular relevant in the study of the dimming regions which are characterized by very low intensities. The reduced pixel resolution of 2.4\arcsec\ is comparable to the horizontal EIS pixel size of 2\arcsec.  
We use the same approach as described in \cite{Vanninathan2018}, to derive for each time step a map of plasma density.
From the retrieved DEM maps, $\phi(T)$, we calculate for each pixel the total emission measure EM and the mean plasma density $n$ using the following expressions:
\begin{equation}
{\rm EM} = \int{ \phi(T)\,\mathrm{d}T} \, , 
\label{eq:density}
\end{equation}
and assuming a filling factor of unity the mean plasma density is calculated as
\begin{equation}
n = \sqrt{\frac{\rm EM}{h}} ,
\label{eq:density}
\end{equation}
with $h$ the integration height contributing to the optically thin emission in the line-of-sight.
In case that the integration height $h$  does not vary considerably during the event, its exact value cancels out when calculating the {\em relative} changes of the density $n$ with respect to the pre-event state, i.e.
\begin{equation}
\frac{n(t)}{n(t_0)} \propto \sqrt{ \frac{ {\rm EM}(t)} { {\rm EM(t_0)} } }. 
\label{eq:dens}
\end{equation}
However, we note that during the event the integration height $h$ might actually change, in particular it may get larger due to the field line stretching and expansion of the plasma in the dimming regions. In this case, the density drop we derive according to Eq. \ref{eq:dens} (i.e.\ under the assumption of constant integration height $h$) provides a lower limit, and the actual density drop corresponding to the measured drop in EM might be larger than the estimate we give.

For the analysis of the time evolution of the plasma parameters in selected regions in the core dimming, the AIA and DEM image sequences were corrected for solar differential rotation, to a reference frame recorded before the event start at 23:06 UT.

\begin{figure*}
\centering
\includegraphics[width=18cm]{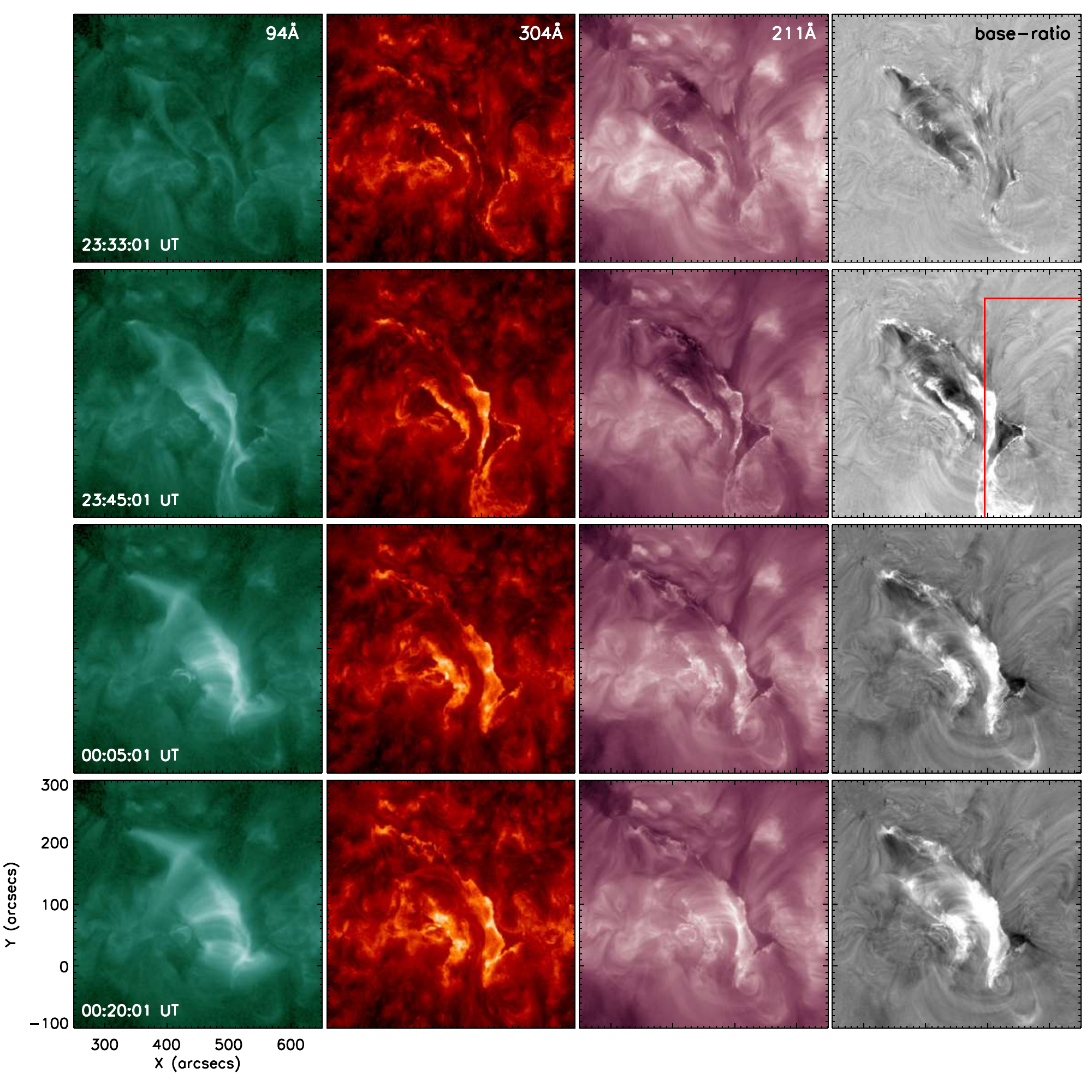}
\caption{Overview of the evolution of the coronal dimming associated with the C7.3 flare and halo CME of 2012 September 27 in SDO/AIA 94 {\AA}, 304 {\AA}, and 211 {\AA} direct as well as 211 {\AA} base ratio images. All images are logarithmically scaled. The red lines drawn in the second AIA 211 {\AA} base ratio panel outlines the Northern and Western borders of the Hinode/EIS raster. The still figure shows snapshots of the event at four time steps (annotated in each panel).  The animated figure online shows the event evolution from 2012 September 27, 23:06 UT to September 28, 00:30 UT. 
}
\label{fig:overview}
\end{figure*}

\begin{figure*}
\centering
\includegraphics[width=17cm]{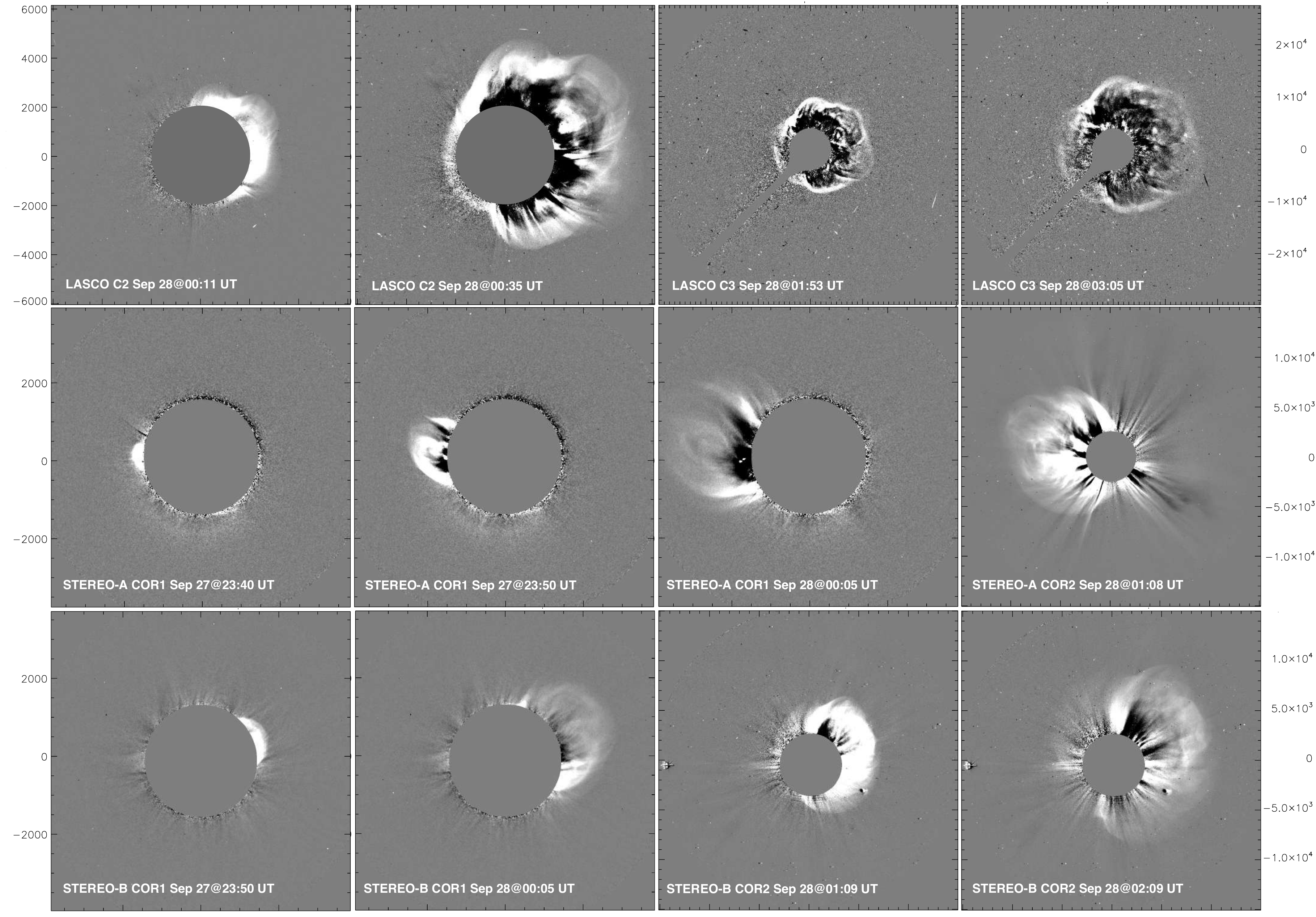}
\caption{Snapshots of the fast halo CME  associated with the C.7 flare observed on 2012 September 28 by SOHO/LASCO C2 and C3 (top panels), STEREO-A (middle panels) and STEREO-B (bottom panels)  COR1 and COR2. The axis scaling is given in arc-seconds, to the left for the inner coronagraphs (C2, COR1) and to the right for the outer coronagraphs (C3, COR2). }
\label{fig:cme}
\end{figure*}

\section{Results}
\label{sec:results}

\subsection{Event overview and magnetic configuration}

Figure \ref{fig:overview} and the associated movie give an overview on the evolution of the event in AIA 94 {\AA} 
(Fe\,{\sc xviii}, $\log T \, [{\rm K}] = 6.8$), AIA 304 {\AA} (He\,{\sc ii}, $\log T \, [{\rm K}] = 4.7$) and  211 {\AA} (Fe\,{\sc xiv}, $\log T \, [{\rm K}] = 6.3$) direct images as well as AIA 211 {\AA} base ratio images. The image sequences illustrate the evolution of the C3.7 two-ribbon flare as well as the formation and evolution of pronounced core dimmings. The hook-shaped ends of the flare ribbons
encompass the dimming regions as is best visible in the 304 {\AA} filter in Figure \ref{fig:overview}. 

In Figure \ref{fig:cme}, we show snapshots of the fast halo CME associated with the event as observed by the LASCO C2 \citep{Brueckner1995} and STEREO COR1 and COR2 \citep{Howard2008} coronagraphs. Notably, LASCO observes the event as a halo CME with a plane-of-sky speed of 950 \kms\, which impacted at Earth about four days later and produced a strong geomagnetic storm. The two STEREO satellites were longitudinally separated from Earth by $\pm105^{\circ}$, and observed the CME on the Eastern and Western limb, respectively.   

As can be seen in Figure \ref{fig:overview}, the core dimming region located at the southern end of the Western flare ribbbon is strongly pronounced in all the AIA filters, indicative of a broad distribution of plasma temperatures. Notably, the second core dimming region located at the Northern end of the Eastern flare ribbon appears most pronounced in the cool AIA 304 {\AA} filter mapping the upper chromosphere and lower transition region plasma, and is barely visible in the coronal AIA filters.  This is an unusual observation. \cite{Dissauer2018b} report that coronal  dimmings appear strongest in the AIA 211 and 193 {\AA} filters,  which are most sensitive to quiet coronal plasma temperatures, and only 15\%  are detectable in the ``cool" 304 {\AA} filter.  
We also note that in all the AIA filters, both core dimmings reveal also elongations along the outer edge  of the two conjugate flare ribbons (cf.\ Figure \ref{fig:overview}, in partcular second row at 23:45 UT). These are associated with the disappearance of the pre-eruptive arcade, as can be most clearly seen in the 211 {\AA} filtergrams (cf.\ the movie accompanying Figure \ref{fig:overview}, 23:06--23:36 UT). 

Inspection of H$\alpha$ filtergrams from the GONG network (not shown) revealed that the event was also associated with the eruption of a long S-shaped filament aligned in North-South direction, with the slow  rise of the filament starting substantially earlier ($\lesssim$22:30 UT) than the flare. The filament eruption can be also seen in the movie associated with Figure \ref{fig:overview}, in particular in the AIA 304 {\AA} and 211 {\AA} filters. 
The movie shows that the core dimming region located at the hooked ends of the Western flare ribbon is located at the Northern footoint of the erupting filament.

\begin{figure*}
\centering
\includegraphics[width=\linewidth]{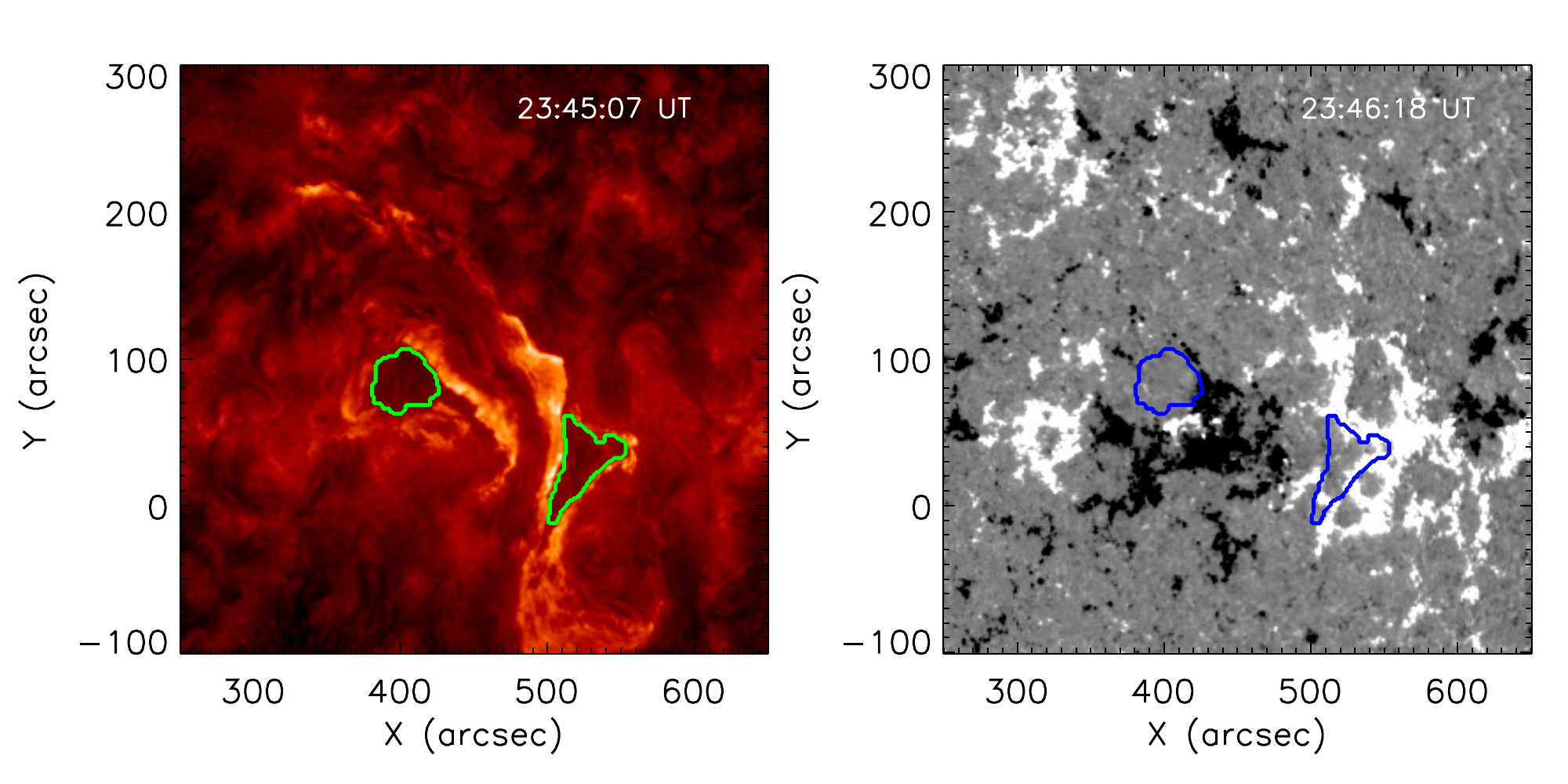}
\caption{\textbf{Left:} SDO/AIA 304\,\AA\ filtergram recorded at 2012 September 27, 23:45 UT. \textbf{Right:} SDO/HMI LOS magnetogram for the same field-of-view scaled to $\pm 50$ G. The contours in both panels outline the borders of the two dimming regions that were segmented from the AIA 304\,\AA\  image. }
\label{fig:hmi}
\end{figure*}

In Figure \ref{fig:hmi}, we show a line-of-sight magnetogram from the Helioseismic Magnetic Imager \cite[HMI;][]{Scherrer2012} together with an AIA 304 {\AA} map at 23:45 UT. The contours of the dimming regions extraced from the AIA 304 {\AA}  filtergram plotted on top of the two images confirm the bipolar nature of the core dimming region, with the Western dimming rooted in positive fields, and the Eastern dimming rooted in negative fields.
In the second panel of the AIA 211 {\AA} images in Figure \ref{fig:overview}, we indicate the boundaries of the FOV of the EIS raster (red lines). As can be seen, the triangular-shaped core dimming region located at the southern end of the Western flare ribbbon was captured by the EIS spectra. It is the main region of interest for the following spectroscopic study.

\begin{figure*}
\centering
\includegraphics[width=\linewidth]{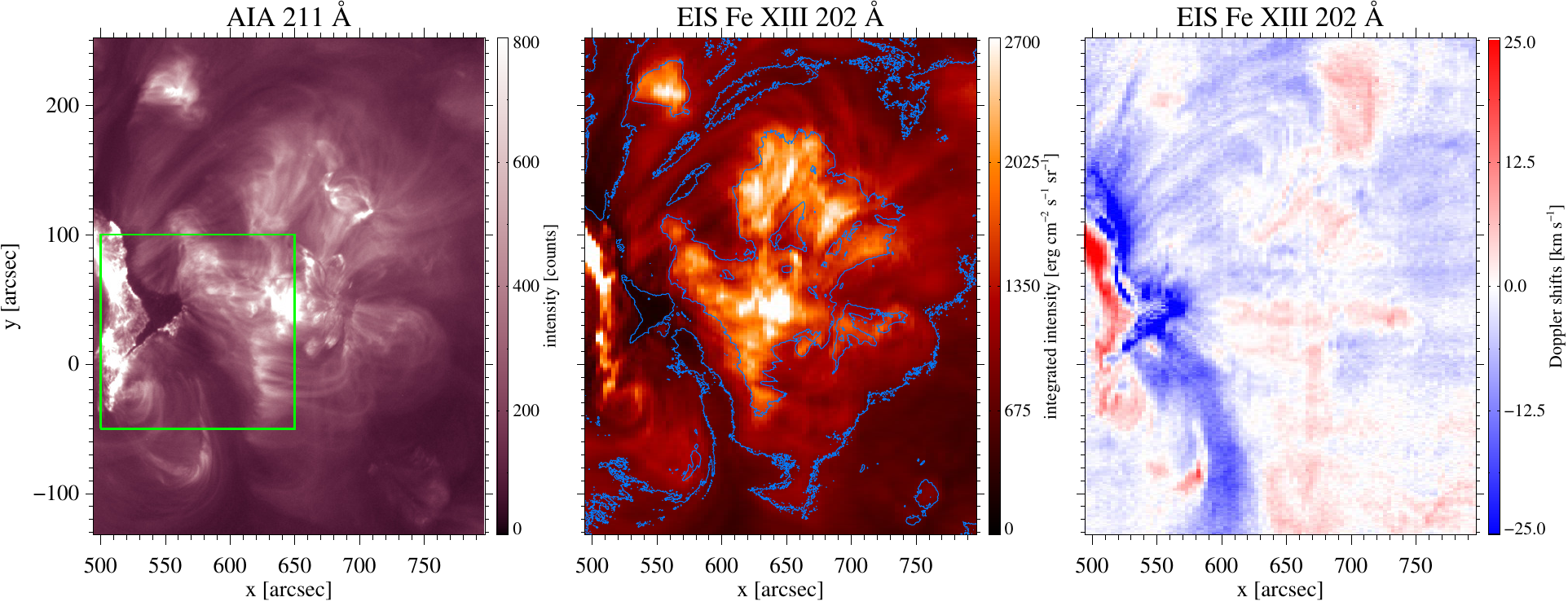}
\caption{\textbf{Left:} AIA 211\,\AA\ image recorded at 2012 September 27, 00:04 UT showing the full EIS field-of-view. The green box represents the zoom of the EIS spectrogram shown in Figure \ref{fig:eis_int_vel}. \textbf{Middle:} Co-aligned EIS intensity map in the Fe\,{\sc xiii} 202.04\,\AA\ spectral line. The blue contours show the AIA 211\,\AA\ intensity levels at 120 and 300 counts, demonstrating the spatial overlap of the structures detected by the EIS and AIA instruments. \textbf{Right:} EIS velocity map corresponding to the intensity map shown in the middle panel. Positive velocities (redshifts; downflows) represent motions toward the solar surface and negative velocities (blueshifts; upflows) represent upward motions. 
}
\label{fig:eis_context}
\end{figure*}

\subsection{EIS spectroscopy}

Figure \ref{fig:eis_context} shows an EIS intensity map in the Fe\,{\sc xiii} 202.04\,\AA\ spectral line (middle panel) and the corresponding velocity map derived from the single Gaussian fits (right panel) together with an AIA 211\,\AA\ image (left panel). The contours plotted on the EIS intensity map are from the AIA 211\,\AA\ image, and demonstrate the spatial overlap of the structures observed by the two instruments after the co-alignment described in Sect. \ref{sec:obs}. 

As core dimming regions are suspected to be footpoints of expanding flux ropes, it is expected to see plasma outflows in this area. This aspect is reinforced by the EIS velocity map shown in Figure\,\ref{fig:eis_context} (right panel). The areas  of the strongest blueshifts on the velocity map (indicative of outflows) coincide with the darkest areas on the intensity map, in particular the dark triangular core dimming region as well as the elongations of the dimming region on the outer front of the flare ribbon. The outflow velocities derived  from the single Gaussian fits  in these regions reveal typical values of some 10 {\kms} but can also reach values up to 40--70 \kms. 
Strong downflows are observed at the location of the flare ribbons, which presumably result from coronal condensation of plasma in the post-flare loops.

Comparison with Figure \ref{fig:overview} and the associated movie shows that the triangular-shaped Western dimming region started to form at about  23:35 UT on 2012 September 27,  and is fully pronounced and dark around 23:45 UT. Note that between 23:35 UT and 23:45 UT, material from the erupting filament partially obscured this dimming region (see AIA 211\,\AA\, difference images), but thereafter it has moved out of the FOV shown in Figure \ref{fig:overview}. After 23:45 UT, we can observe that the dimming region is still slightly growing toward the Western direction until $\sim$00:05 UT on 2012 September 28. This growth is most distinctly seen in the evolution of the hooked  flare ribbon that forms the Western boundary of the dimming region, which we interpret as a signature of magnetic reconnection along the boundary of the newly ``opened" fields. 

When studying the EIS intensity and velocity maps shown in Figure \ref{fig:eis_context} in relation to the evolution observed in AIA, it is important to note that the EIS slit rastered the FOV 
from West (start 23:10 UT) to East (end 00:21 UT).
This means that the triangular-shaped dimming region (starting at  its Western-most segment at $x \sim 570\arcsec)$ is the first time covered by the EIS slit at 
$\sim$00:03 UT on 2012 September 28.

\begin{figure}
\centering
\includegraphics[width=\linewidth]{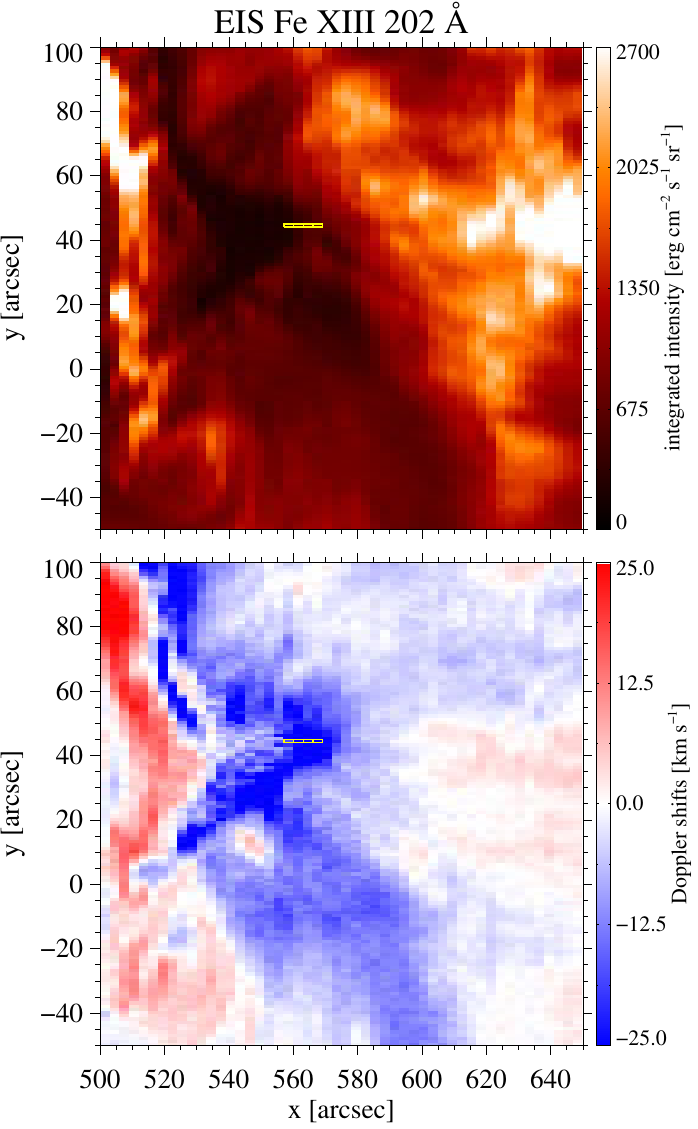}
\caption{EIS intensity (top) and velocity (bottom) maps derived from the Fe\,{\sc xiii} 202.04\,\AA\ spectral line showing a zoom to the dimming region under study (cf.\ green box in Figure \ref{fig:eis_context}). The yellow quadrangles mark pixels which correspond to the EIS spectral profiles shown in Figure \ref{fig:eis_fexiii_profiles}.}
\label{fig:eis_int_vel}
\end{figure}

\begin{figure*}
\centering
\includegraphics[width=\linewidth]{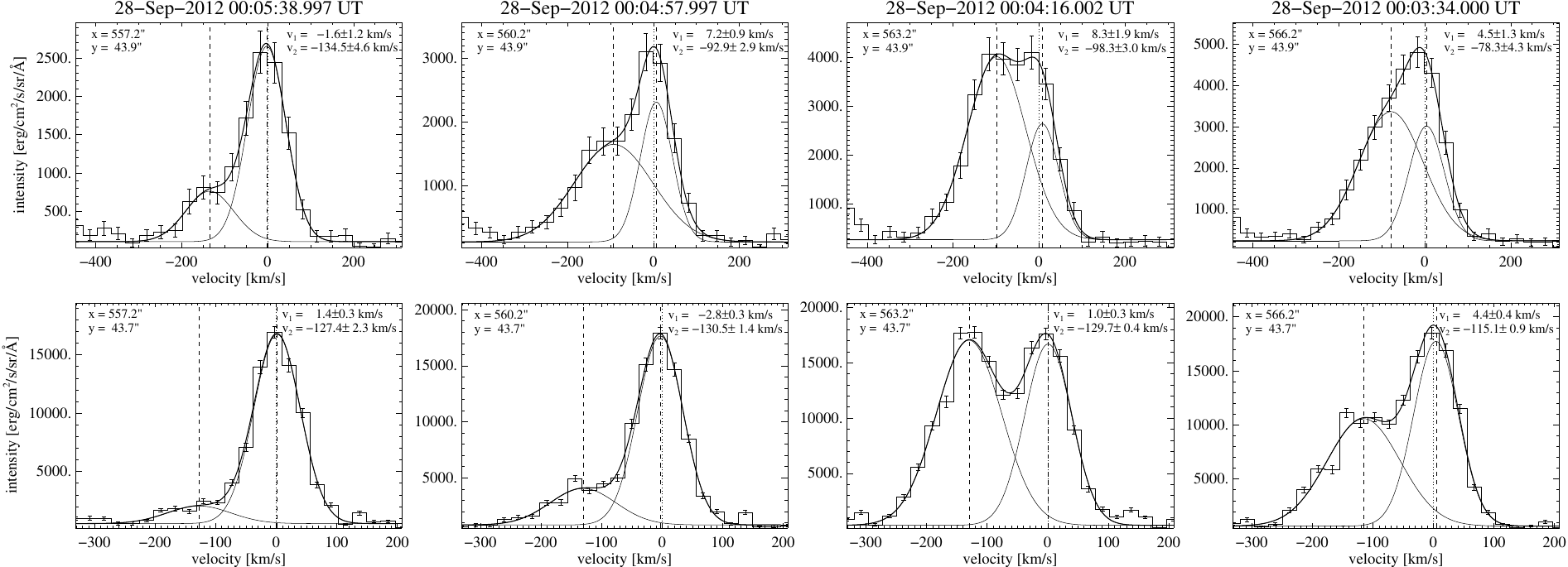}
\caption{EIS spectra in selected pixels in the dimming region. Fe\,{\sc xiii} 202.04\,\AA\  (upper row) and Fe\,{\sc xv}  284.16\,\AA\ (lower row) line profiles together with the double Gaussian fits for the EIS pixels marked in Figure\,\ref{fig:eis_int_vel} . 
The spatial coordinates of the displayed spectral profiles are shown in the upper left corners of the sub-panels. The velocities estimated for both spectral components are displayed in the upper right corners.}
\label{fig:eis_fexiii_profiles}
\end{figure*}

Figure\,\ref{fig:eis_int_vel} shows a zoom into the EIS Fe\,{\sc xiii} 202.04\,\AA\ intensity and velocity maps, where we inspected the spectral line profiles of the dimming regions. 
Particular pixels of interest, for which we show the spectral line profiles in Figure\,\ref{fig:eis_fexiii_profiles}, are indicated by yellow quadrangles. The top panels of  Figure\,\ref{fig:eis_fexiii_profiles} shows four spectral profiles in the Fe\,{\sc xiii} 202.04\,\AA\ line, the bottom panel shows for the same pixels the line profiles in the Fe\,{\sc xv} line at 284.16\,{\AA}, both together with the corresponding double-Gaussian fits. The four pixels shown have been covered by the EIS slit at different times, in total about 2 minutes apart. The rightmost pixel was on the EIS slit 
at about 00:03:30 UT, and the leftmost pixel at about 00:05:40 UT. 

The EIS Fe\,{\sc xiii} and Fe\,{\sc xv} line profiles reveal an interesting behavior. Most of the profiles within the dimming region can be fitted by a single Gaussian. However, as we go towards the regions of strongest blue-shifts, we see that a second component starts to develop. All the spectra shown in Figure\,\ref{fig:eis_fexiii_profiles} reveal a clear double component, and can be well fitted by two Gaussians. The primary component is consistent with 
nearly stationary plasma; the derived speeds in this component are all very small, ranging from $-3$\,\kms\ to  $+8$\,\kms .

The second component reveals upflowing plasma  with high speeds. The velocities  obtained from this component range from $-78$\,\kms\ to $-135$\,\kms\ for the Fe\,{\sc xiii} 202.04\,\AA\ line, and $-115$\,\kms\ to $-130$\,\kms\ for the Fe\,{\sc xv} 284.16\,{\AA} line. Note that in several pixels, the intensity of the second (i.e., upflowing) component matches that of the primary (i.e., stationary) component, and may even overshoot it by almost a factor of two. Finally, we note that also in the Si\,{\sc vii} 275.37\,\AA\ line we identified several pixels in the dimming region that show a double component indicative of upflows with velocities of $-60$ to  $-70$\,\kms. In general, the Si\,{\sc vii} data is very noisy making it difficult to obtain line profiles in every pixel.

\subsection{Density diagnostics}
\label{ssec:density}

To estimate the electron densities in the core dimming region under study, we use the intensity ratio of the Fe\,{\sc xiii} spectral lines observed at 196.55\,\AA\ and 202.04\,\AA. 
We note that the  presence of blends in other EIS spectral lines, including the strongest lines of Fe\,{\sc xii} at 195.12\,\AA\ and Fe\,{\sc xiii} at 203.82\,\AA, renders them complicated to be used for our study. Correcting the blends and then fitting double component profiles requires more than three Gaussians, which reduces the accuracy of the results obtained.

As is seen in the spectral line profiles plotted in Figure\,\ref{fig:eis_fexiii_profiles}, the EIS spectra in these areas of the core dimming region consist of two components indicative of a static and a fast upflowing plasma. Two-component fitting of both spectral lines was performed in order to separate the emissions from the outflowing and the stationary plasma. The best separation between the two-components is obtained for the position ($x=563.2$\arcsec, $y=43.9$\arcsec) in the core dimming region. The calculated electron densities for this EIS pixel is $2.25\times 10^{9}$\,cm$^{-3}$ for the outflowing plasma and $1.13\times 10^{9}$\,cm$^{-3}$ for the stationary plasma.

We note that the Fe\,{\sc xiii} line at 196.55\,\AA\  used for density diagnostics is very close to the Fe\,{\sc xii} line at 196.64 \,\AA\, which forms at similar temperatures. Thus, the Fe\,{\sc xii} 196.64 line also shows an outflow component which lies close to the stationary component of the Fe\,{\sc xiii} 196.55 line. This affects the density diagnostics for the stationary component. (The density diagnostics for the outflow component is likely not affected.) Therefore, we were very careful with the interpretation of the determined densities, and we give only the values derived for the one EIS pixel where we identified the strongest upflows. It was the only spectrum where we could fit both lines with two components, so that we had some estimate how much the dynamic component of the Fe\,{\sc xii} line is affecting the static component of the Fe\,{\sc xiii} line.

\begin{figure}
\centering
\includegraphics[width=8.5cm]{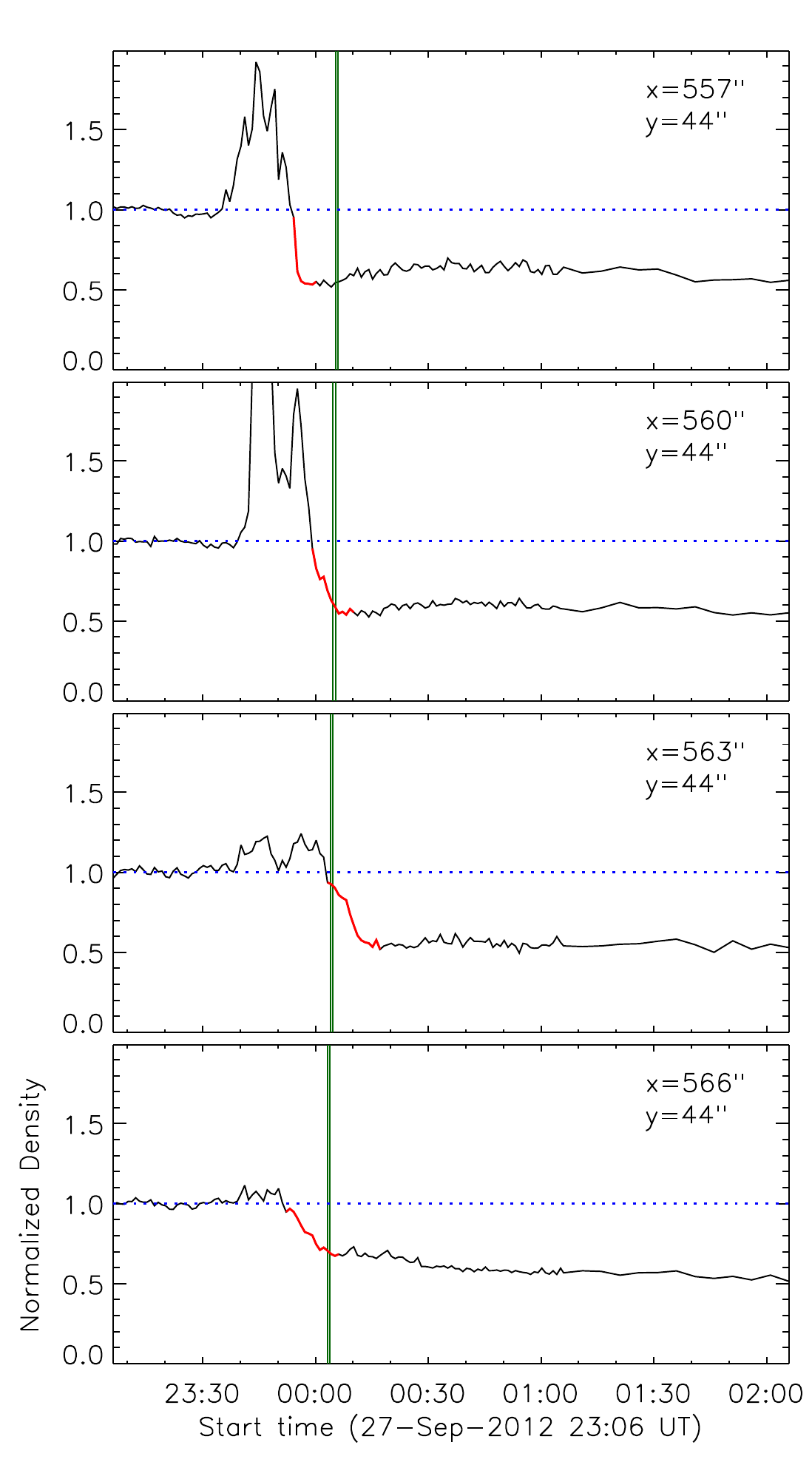}
\caption{Time evolution of the relative density changes, i.e.\ actual density divided by its pre-event value, derived from AIA DEM maps for pixels corresponding to the EIS spectra in Figure\,\ref{fig:eis_fexiii_profiles}. The segments in red indicate the phase of the impulsive density drop: Its start is defined by the data steeply decreasing below 1, and its end is calculated from the time derivative of the curve (defined by the first positive increase). The green vertical lines indicate the time at which the corresponding pixel was covered by the EIS slit.}
\label{fig:DEM_parameters}
\end{figure}

As the EIS data consists of only one raster, meaning only one observation per EIS pixel, we use AIA data to study the temporal evolution of the dimming region and the associated density depletion from the reconstructed DEM maps. We have used 3.5\,hrs of AIA data, which includes 30\,min of pre-flare to estimate a background level, and continues until 3 hours after the start of the flare. 

In Figure \ref{fig:DEM_parameters}, we show the time evolution of the plasma density derived in the regions corresponding to the EIS pixels shown in Figures \ref{fig:eis_int_vel} and \ref{fig:eis_fexiii_profiles}. These values are normalized to the pre-event state, so we can directly assess the percentage density drop within the dimming regions. In the pixels shown, the density drops by 40--50\%, and remains at these reduced levels for the whole period shown.
The distinct (double) density enhancements observed {\em before} the density drop related to the coronal dimming are caused by the erupting filament plasma and the bright ribbbon fronts on the outer border of the growing dimmig region that come into the pixel's FOV.

\begin{figure}
\centering
\includegraphics[width=8cm]{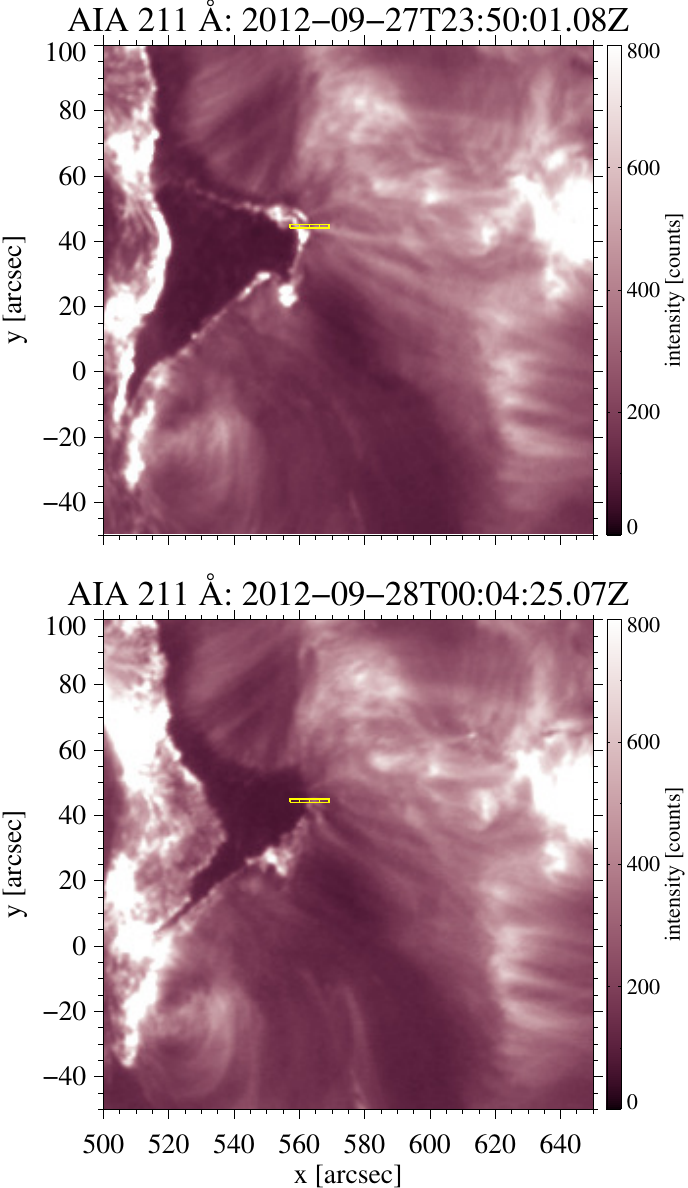}
\caption{Evolution of the Western dimming region in SDO/AIA 211 {\AA} filtergrams. We show the same zoomed FOV as for the EIS maps in Figure  \ref{fig:eis_int_vel}. The yellow boxes mark pixels for which we show the EIS spectral profiles in Figure \ref{fig:eis_fexiii_profiles}. The still figure shows the event at two time steps (annotated in each panel).
The animated figure online shows the event evolution from 2012 September 27, 23:50 UT to September 30, 01:16 UT.  
}
\label{fig:aia_evolution}
\end{figure}

To relate the evolution of the densities reconstructed from the AIA DEM maps (Figure \ref{fig:DEM_parameters}) in the 4 pixels indicated in Figure  \ref{fig:eis_int_vel} with the corresponding spectra and flows determined from the EIS scan (Figure \ref{fig:eis_fexiii_profiles}), we need to relate the dynamics observed in AIA with the time at which the EIS slit covered the particular pixels. We first discuss the behavior of the top three pixels plotted in Figure \ref{fig:DEM_parameters}, and thereafter the bottom-most one which is related to different structures as we will show.

In the density evolution retrieved from the DEM maps (Figure \ref{fig:DEM_parameters}), we can see that the sharp density drop due to the dimming occurs first in the left-most (Eastern-most) pixel, and thereafter in the two neighboring pixels to the right. This is consistent with the expansion of the dimming region toward West, i.e.\ from left to right (see Figure \ref{fig:aia_evolution} and the associated movie). 
On the other hand, EIS was in scanning mode 
and we may expect to see the largest upflow components in those pixels which have been activated as dimming pixels (i.e.\ the associated fields were ``opened") closest in time when the EIS slit was mapping them. 

The EIS scan was from West to East,
and the strongest upflow component is observed at the pixel  $(x=563\arcsec, y=43.9\arcsec)$, which was covered by the slit at 00:04:16 UT. This is the pixel where we find that the total amount of upflowing plasma is larger than the stationary plasma component. The AIA density evolution shows that in this pixel the strong drop in the density, indicative of the impulsive dimming onset, starts {\em exactly at that time} when it was mapped by the EIS slit. In the neighboring EIS pixel to the left, centered at  $x=560 \arcsec$, the impulsive density drop occurred during 23:58 to 00:06 UT, and the EIS slit covered it {\em during} this time range at 00:04:58 UT. This finding is consistent with  EIS  still showing a large upflow component, but less strong than in the previous pixel. Finally, the pixel centered  at  $x=557\arcsec$ reveals its impulsive density drop during 23:53 to 23:57 UT, i.e.\ {\em before} it was covered  by  EIS  at 00:05:39 UT, which is consistent with the even smaller upflow plasma component observed here.  

The EIS pixel centered at  $x=566 \arcsec$ is an exception to the timing behavior described above, as its impulsive density drop occurs already during 23:55 to 00:04 UT, i.e. before the westward growing dimming bordered by the bright flare ribbon reached this pixel. Inspection of the movie and still images in Figure \ref{fig:aia_evolution} reveals that this region dims due to the disappearance of a loop connecting toward South-West direction. However, also this pixel follows the overall rule to show a significant upflowing plasma component as it was covered by the EIS slit {\em during} the phase of impulsive density drop (similar to pixel $x=560 \arcsec$).

Finally, we note that the Western dimming region under study was formed in most of its extent already at 23:45 UT on 2012 September 27. The strongest upflow components in the dimming region that we measure in the EIS scan refer to the period 00:03 to 00:05 UT on 2012 September 28 when the EIS slit started to actually scan the dimming region. This means we missed the formation and early evolution of the major part of the coronal dimming in the EIS spectra. However, even at these later times,  distinct upflows are registered by EIS all over the dimming region (cf.\ Figure \ref{fig:hmi}, right panel). The clear two-component spectra, indicative of strong upflowing plasma components, are observed predominantly for those pixels that are located on the {\em newly formed} outer dimming segments, behind the propagating bright flare ribbon that surrounds the dimming region. 
%We may speculate that mapping the whole dimming region with a spectrometer about 20  min earlier, we might have measured strong upflow components over larger areas of the dimming region.

\begin{figure}
\centering
\includegraphics[width=8.5cm]{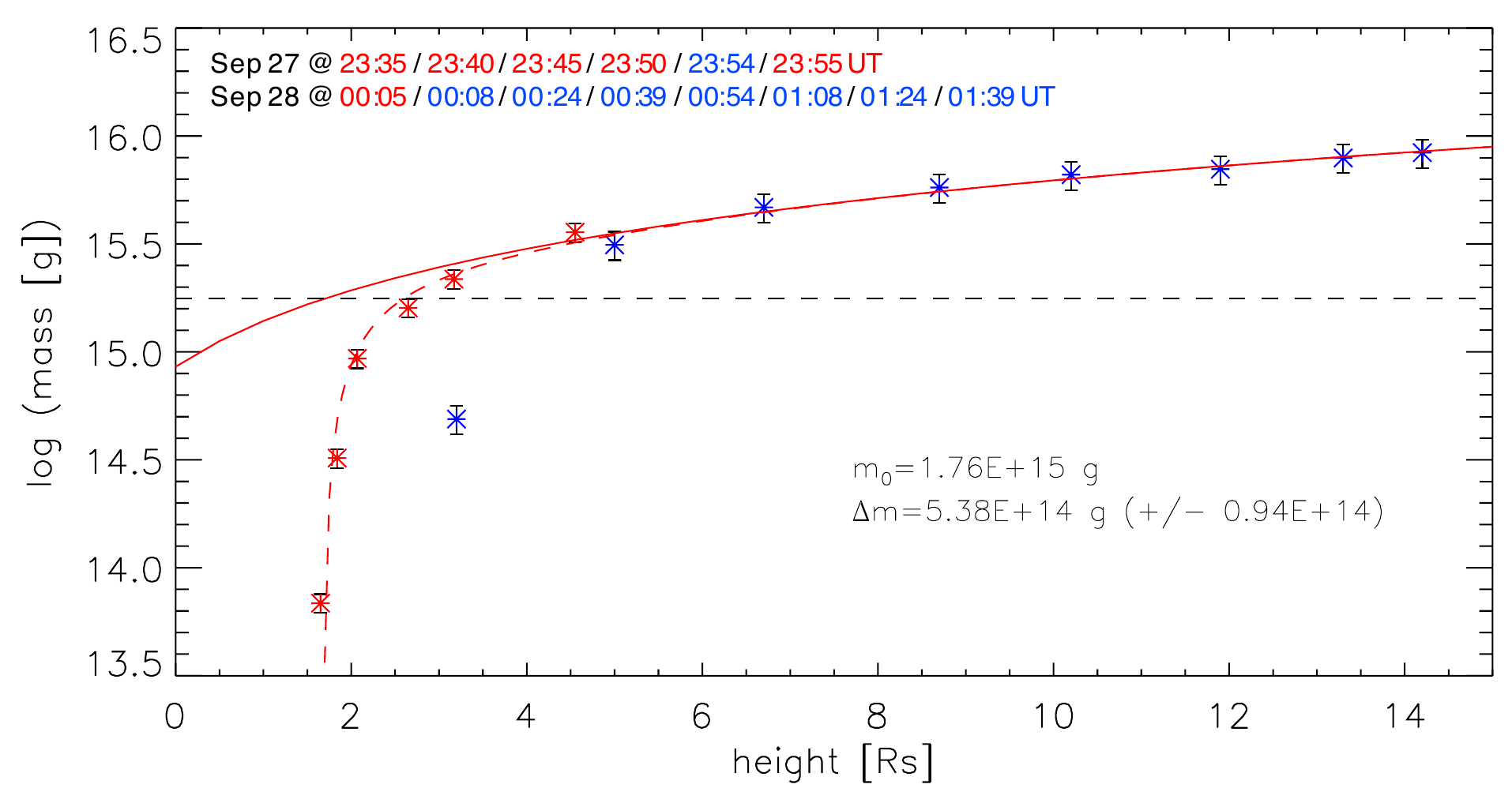}
\caption{Mass evolution against height for the CME associated with the C3.7 flare of 2012 September 27, derived from STEREO-A COR1 (red asterisks) and COR2 (blue asterisks) white-light observations. The observing times related to the data points are given in the inset.
The dashed red line gives the fit to the COR1 data, the solid red line shows the real mass evolution after accounting for the occulter effect. The derived values for the mass increase per solar radius ($\Delta$m), the inital CME  mass $m_0$ 
are given in the legend. The dashed line indicates  the inital CME  mass $m_0$.}
\label{fig:cme-mass}
\end{figure}

\subsection{Mass evolution of the associated CME}

In the following, we study the mass evolution of the associated white-light CME and compare the outcome to the plasma outflows from the coronal dimming regions. The CME was observed by the coronagraphs onbard three satellites located at different positions in the heliosphere, SOHO, STEREO-A and STEREO-B (cf.\ Figure \ref{fig:cme}). For the CME analysis, we use the COR1 and COR2 coronagraphs onboard STEREO-A, which observed the CME close to the limb, i.e.\ in its plane-of-sky (for STEREO-A, the source region was located at 72$^\circ$ East). The kinematics of the CME front derived from STEREO-A gives a mean speed of 1250 \kms. This is higher than the 950 {\kms} listed in the LASCO CME catalogue, which is to be expected, as LASCO observed the CME head-on as a halo and thus the LASCO measurements mostly reflect the CME expansion and not its propagation speed. 

We use the method of \cite{Bein2013} to derive the mass evolution from the STEREO-A coronagraph data. This method accounts for the effect of the occulter disk, which masks part of the CME structure in the low corona, in detail depending on the CME direction and 3D structure relative to the viewing direction of the observing spacecraft. From this information, an effective occultation height is derived, and  is used to separate this ``geometric" effect on the derived mass evolution from the physical one.

Figure \ref{fig:cme-mass} shows the evolution of the CME mass calculated with this method as a function of height of the CME front. The steep mass increase in the COR1 field-of-view (1.4--4 R$_s$) is due 
to mass emerging from behind the occulting disk when the CME moves outward, i.e.\ is not a  real mass increase. A similar trend is seen for the early evolution of CME mass in COR2. % COR2 covers in total a height range from 2.5 to 15 R$_s$. 
We apply the functional fit developed in \cite{Bein2013} to quantify both the effect of the coronagraph occulter low in the corona and the actual mass evolution. From the fit (shown by the solid red line in Figure \ref{fig:cme-mass}), we can derive the effective occultation height, the initial CME mass injected ($m_0$), and a mean rate of increase of CME mass  ($\Delta m$). For the event under study, we obtain a mass increase rate of $(5.4 \pm 0.9) \times 10^{14}$ g per solar radius passed by the CME. For a mean CME speed of 1250 \kms, this corresponds to a mass increase  rate of about 
$(9.7 \pm 1.6) \times 10^{11}$~g~s$^{-1}$.

From the EUV spectral and imaging information provided by EIS and AIA, we can obtain an order-of-magnitude estimate of the mass flux from the dimming outflows. The total area of the dimming regions outlined in Figure \ref{fig:hmi} is  $A_{\rm d} = 1.6\times 10^{19}$ cm$^{2}$. The upflow speeds in the dimming pixels which reveal a clear double Gaussian component are of the order of $v_{\rm up}\approx 100$ \kms. The number density in the upflowing plasma derived from EIS spectroscopy is $n_{\rm up} = 2.25 \times 10^9$~cm$^{-3}$, which relates to a mass density of  $\rho_{\rm up} = 4.8 \times 10^{-15}$~g~cm$^{-3}$ using a mean molecular mass for solar abundances of $\mu=1.27$ \citep{Anders1989}. From these quantities, we can estimate the mass loss rate related to the upflowing plasma in the dimming region as 
\begin{equation}
\frac{dm_{\rm up}}{dt} = A_{\rm d} \; \rho_{\rm up}  \; v_{\rm up} = 7.7 \times 10^{11} {\rm \; g \; s}^{-1} \, .
\end{equation}
We note that this should probably be considered as an upper estimate, as we assumed the whole dimming region to contribute with the same ``strength" (in terms of density of the upflowing plasma and outflow speed) as derived from the EIS pixels that showed the clearest and strongest upflow components. Also, we do not know how long the outflows are actually present as we have only one EIS scan. However, it is interesting to note that this order of magnitude estimate of the mass loss rate in the plasma upflows calculated from the coronal dimming regions is of the same magnitude as the mass increase rate observed in the white-light CME.

\section{Discussion and Conclusions}

In this paper, we studied a coronal dimming caused by the launch of a fast halo CME (deprojected speed of 1250 \kms) 
associated with the C3.7 two-ribbon flare on 2012 September 27, using Hinode/EIS spectroscopy and SDO/AIA DEM analysis. Our main findings and conclusions are the following:
\begin{itemize}
\item We observe bipolar core dimming regions encompassed by hook-shaped flare ribbons located at the ends of the flare-related polarity inversion line (Figure \ref{fig:hmi}). The event is associated with the eruption of a filament, which reveals its footpoints at the core dimming regions formed during its ejection (Figure \ref{fig:overview}). The dimming region grows in the regions directly behind the expanding hooked parts of the bright flare ribbons.
%, indicative of magnetic reconnection between the erupting flux rope and the surrounding corona.  
In addition, the core dimming regions reveal also elongations along the outer edges of the two conjugate flare ribbons. 
\item Plasma upflows with typical speeds of some 10 {\kms} but up to 40--70 {\kms} (as derived from single Gaussian fits) are observed all over the Western core dimming region covered by Hinode/EIS, although parts of it were mapped by the EIS slit $\gtrsim$20 min {\em after} the local dimming formation
(Figures \ref{fig:eis_context}, \ref{fig:eis_int_vel}).
\item Distinct double component spectra  indicative of the superposition of a stationary and a fast upflowing plasma component are observed at those pixels at the growing dimming border, which were mapped by the EIS slit close in time (i.e.\ within a few minutes) of their impulsive dimming onset (Figure \ref{fig:eis_fexiii_profiles}). In some EIS pixels, the upflowing plasma component is even dominant over the stationary one (by up to a factor of 2). 
%
%\item 
The velocities derived for the upflowing component are up to $-130$\,\kms\ for the Fe\,{\sc xiii} 202.04\,\AA\ ($\log T \, [{\rm K}]  = 6.2$) and Fe\,{\sc xv} 284.16\,{\AA} ($\log T \, [{\rm K}]  = 6.3$) lines, and up to $-70$\,\kms\ in the Si\,{\sc vii} 275.37\,\AA\ line ($\log T \, [{\rm K}]  = 5.8$).
\item 
Electron density estimates in the core dimming region using the intensity ratio of the EIS Fe\,{\sc xiii} 196.55\,\AA\ and 202.04\,\AA\ spectral lines give
$2.25\times 10^{9}$\,cm$^{-3}$ for the outflowing plasma and $1.13\times 10^{9}$\,cm$^{-3}$ for the stationary plasma.
The dynamics of the relative density changes in core dimming regions derived from SDO/AIA DEM analysis reveals impulsive reductions by 40--50\% within $\lesssim$10 min. The densities remain at this reduced levels for $>$2.5 hours after their initiation (Figure \ref{fig:DEM_parameters}).
\item From the EIS dimming observations, we estimate a mass outflow from the dimming region of about $8 \times 10^{11} {\rm \; g \; s}^{-1}$. This estimate is close to the mass increase of about $1 \times 10^{12} {\rm \; g \; s}^{-1}$ observed in the associated white light CME in the STEREO COR1 and COR2 fields (Figure \ref{fig:cme-mass}).
\end{itemize}

The elongated dimmings on the outer front of the flare ribbons can be understood in the frame of the standard 2.5D eruptive flare model, in which the overlying coronal arcade is stretched by the rising CME causing  dimmings at its footpoints \citep{Forbes2000,Cheng2016}. Subsequently, the arcade field then reconnects and causes the bright flare ribbons at the edge of the elongated dimming regions. In this case it is expected that the dimming appears {\em before} the flare ribbons. In the event under study, we can in fact see how the pre-eruptive arcade associated with the elongated dimmings disappears while subsequently the dimming and the flare ribbons form (cf.\ the movie accompanying Figure \ref{fig:overview}, in particular the 211 {\AA} data).

However, to understand the morphology of the core dimming regions encompassed by the hook-shaped (double J-shaped) flare ribbons, and the relation between the local flare ribbon expansion and the growth of the encompassed dimming region needs the framework of three-dimensional extensions to the standard eruptive flare model \cite[][]{Aulanier2012,Janvier2015}. 
In this three-dimensional geometry, the flux rope is surrounded by quasi-separatrix layers (QSLs), i.e.\ thin layers of steep gradients in field-line connectivities, and 
the QSL footprints have a curved shape whose endings map to the outer edges of the flux-rope footpoints, which can be related with the hook-shaped flare ribbons \cite[][]{Demoulin1996,Aulanier2019}.

The double component Hinode/EIS spectra indicating the superposition of stationary and upflowing plasma components in core dimming regions, and outflows with velocities up to $130$ {\kms} support the findings of \cite{Tian2012} who performed an extensive spectroscopic study of several well observed dimming events. In our case, we find that in pixels which are initiated as dimmings close in time to their mapping by the scanning EIS slit, the outflow components can be of the same order (and even bigger) than the stationary plasma component, whereas in \cite{Tian2012} the outflow components observed are mostly weak (of the order of 10\% of the stationary component).  

The strong outflow component we find in the coronal EIS spectra is consistent with the distinct and impulsive density reductions by 40--50\% derived from the SDO/AIA DEM analysis. Similar and even stronger reductions of 50--70\% in core dimming regions derived from DEM analysis have been reported in the six events studied in \cite{Vanninathan2018}, suggesting that about half up to two-third of the pre-eruption plasma has escaped the core dimming regions.
We note that these values are most probably lower estimates, as accounting for an increase of the integration height $h$ during the event (which may be expected due to the field line stretching and plasma expansion in the dimming regions), would result in even larger density drops calculated from the EM evolution.

We note that there are basically two ways to look at the upflows observed in the dimming regions. Either they are interpreted as mass outflows in the regions where the magnetic field is locally opened and/or expanding due to the erupting CME \cite[e.g.,][]{Harrison2000,Harra2007} or they are interpreted as a source of plasma refilling the corona following an eruption \cite[][]{Tian2012}.  In the case under study, we strongly favor the first option for the following reasons: the high upflow speeds derived, the strong upflow component which is of the same order as the stationary plasma component, and the impulsive reduction of density in the dimming regions associated with strong upflows which remains at these low levels for hours after the eruption without indication for refill. The order of magnitude estimates of the mass loss rate by the plasma flows in the dimming region,  which lies in the same range as the mass increase rate in the associated white-light CME, also supports this interpretation. 

\cite{Bein2013} studied the mass evolution in the coronagraphic field-of-view for a set of 25 CMEs. They found a distinct mass increase (of the order of a few percent) in all of the cases, and that the CME center of mass reveals a tendency to fall backward with respect to the overall outward moving CME structure. These findings also favor the interpretation that the source of the mass increase are plasma flows from below and not pile-up of coronal material ahead of the CME \cite[][]{Vourlidas2010,Feng2015,Howard2018}. A similar 
conclusion was also drawn in the case study by \cite{Temmer2017} from comparison of the area evolution of the core dimming regions and the mass increase in the white-light CME.

Finally, we note that the current case study reinforces the extensive potential of coronal dimmings to better understand and characterize solar eruptions. 
Coronal dimmings contain important information on the pre-eruptive configuration of CMEs, their initiation and early impulsive evolution, the relation to the associated flare as well as on some of the CME's key properties like mass and speed \cite[see also the recent statistical studies by][]{Dissauer2018b,Dissauer2019}.

\label{sec:concl}

\acknowledgments

SDO data are courtesy of the NASA/SDO AIA and HMI science teams. Hinode is a Japanese mission developed and launched by ISAS/JAXA, with NAOJ as domestic partner and NASA and UKSA as international partners. It is operated by these agencies in co-operation with ESA and NSC (Norway). SOHO is a project of international cooperation between ESA and NASA. The STEREO/SECCHI data are prepared by an international consortium of NASA Goddard Space Flight Center (USA), Lockheed Martin Solar and Astrophysics Lab (USA), Naval Research Laboratory (USA), Rutherford Appleton Laboratory (UK), University of Birmingham (UK), Max-Planck-Institut f\"ur Sonnensystemforschung (Germany), Institut d'Optique Theorique et Appliquee (France), Institut d'Astrophysique Spatiale (France) and Centre Spatiale de Liege (Belgium).
This work was supported by the project of the \"{O}sterreichischer Austauschdienst (OeAD) and the Slovak Research and Development Agency (SRDA) under grant Nos. SK 01/2016 and SK-AT-2017-0009. AMV, KD, MT and KV gratefully acknowledge the support by the Austrian Science Fund (FWF): P27292-N20, and the Austrian Space Applications Programme of the Austrian Research Promotion Agency FFG (ASAP-11 4900217, ASAP-14 865972). P.G. acknowledges support by the project VEGA 2/0004/16.

%\clearpage
%\bibliography{references}

\begin{thebibliography}{}
\expandafter\ifx\csname natexlab\endcsname\relax\def\natexlab#1{#1}\fi
\providecommand{\url}[1]{\href{#1}{#1}}

\bibitem[{{Anders} \& {Grevesse}(1989)}]{Anders1989}
{Anders}, E., \& {Grevesse}, N. 1989, \gca, 53, 197

\bibitem[{{Attrill} {et~al.}(2009){Attrill}, {Engell}, {Wills-Davey}, {Grigis},
  \& {Testa}}]{Attrill2009}
{Attrill}, G.~D.~R., {Engell}, A.~J., {Wills-Davey}, M.~J., {Grigis}, P., \&
  {Testa}, P. 2009, \apj, 704, 1296

\bibitem[{{Aulanier} \& {Dud{\'{\i}}k}(2019)}]{Aulanier2019}
{Aulanier}, G., \& {Dud{\'{\i}}k}, J. 2019, \aap, 621, A72

\bibitem[{{Aulanier} {et~al.}(2012){Aulanier}, {Janvier}, \&
  {Schmieder}}]{Aulanier2012}
{Aulanier}, G., {Janvier}, M., \& {Schmieder}, B. 2012, \aap, 543, A110

\bibitem[{{Bein} {et~al.}(2013){Bein}, {Temmer}, {Vourlidas}, {Veronig}, \&
  {Utz}}]{Bein2013}
{Bein}, B.~M., {Temmer}, M., {Vourlidas}, A., {Veronig}, A.~M., \& {Utz}, D.
  2013, \apj, 768, 31

\bibitem[{{Brueckner} {et~al.}(1995){Brueckner}, {Howard}, {Koomen},
  {Korendyke}, {Michels}, {Moses}, {Socker}, {Dere}, {Lamy}, {Llebaria},
  {Bout}, {Schwenn}, {Simnett}, {Bedford}, \& {Eyles}}]{Brueckner1995}
{Brueckner}, G.~E., {Howard}, R.~A., {Koomen}, M.~J., {et~al.} 1995, \solphys,
  162, 357

\bibitem[{{Cheng} \& {Qiu}(2016)}]{Cheng2016}
{Cheng}, J.~X., \& {Qiu}, J. 2016, \apj, 825, 37

\bibitem[{{Culhane} {et~al.}(2007){Culhane}, {Harra}, {James}, {Al-Janabi},
  {Bradley}, {Chaudry}, {Rees}, {Tandy}, {Thomas}, {Whillock}, {Winter},
  {Doschek}, {Korendyke}, {Brown}, {Myers}, {Mariska}, {Seely}, {Lang}, {Kent},
  {Shaughnessy}, {Young}, {Simnett}, {Castelli}, {Mahmoud}, {Mapson-Menard},
  {Probyn}, {Thomas}, {Davila}, {Dere}, {Windt}, {Shea}, {Hagood}, {Moye},
  {Hara}, {Watanabe}, {Matsuzaki}, {Kosugi}, {Hansteen}, \&
  {Wikstol}}]{Culhane2007}
{Culhane}, J.~L., {Harra}, L.~K., {James}, A.~M., {et~al.} 2007, \solphys, 243,
  19

\bibitem[{{DeForest} {et~al.}(2013){DeForest}, {Howard}, \&
  {McComas}}]{DeForest2013}
{DeForest}, C.~E., {Howard}, T.~A., \& {McComas}, D.~J. 2013, \apj, 769, 43

\bibitem[{{D{\'e}moulin} {et~al.}(1996){D{\'e}moulin}, {Priest}, \&
  {Lonie}}]{Demoulin1996}
{D{\'e}moulin}, P., {Priest}, E.~R., \& {Lonie}, D.~P. 1996, \jgr, 101, 7631

\bibitem[{{Dissauer} {et~al.}(2019){Dissauer}, {Veronig}, {Temmer}, \&
  {Podladchikova}}]{Dissauer2019}
{Dissauer}, K., {Veronig}, A.~M., {Temmer}, M., \& {Podladchikova}, T. 2019,
  \apj, 874, 123

\bibitem[{{Dissauer} {et~al.}(2018{\natexlab{a}}){Dissauer}, {Veronig},
  {Temmer}, {Podladchikova}, \& {Vanninathan}}]{Dissauer2018a}
{Dissauer}, K., {Veronig}, A.~M., {Temmer}, M., {Podladchikova}, T., \&
  {Vanninathan}, K. 2018{\natexlab{a}}, \apj, 855, 137

\bibitem[{{Dissauer} {et~al.}(2018{\natexlab{b}}){Dissauer}, {Veronig},
  {Temmer}, {Podladchikova}, \& {Vanninathan}}]{Dissauer2018b}
---. 2018{\natexlab{b}}, \apj, 863, 169

\bibitem[{{Feng} {et~al.}(2015){Feng}, {Wang}, {Shen}, {Shen}, {Inhester},
  {Lu}, \& {Gan}}]{Feng2015}
{Feng}, L., {Wang}, Y., {Shen}, F., {et~al.} 2015, \apj, 812, 70

\bibitem[{{Forbes} \& {Lin}(2000)}]{Forbes2000}
{Forbes}, T.~G., \& {Lin}, J. 2000, Journal of Atmospheric and
  Solar-Terrestrial Physics, 62, 1499

\bibitem[{{Hannah} \& {Kontar}(2012)}]{Hannah2012}
{Hannah}, I.~G., \& {Kontar}, E.~P. 2012, \aap, 539, A146

\bibitem[{{Hannah} \& {Kontar}(2013)}]{Hannah2013}
---. 2013, \aap, 553, A10

\bibitem[{{Hansen} {et~al.}(1974){Hansen}, {Garcia}, {Hansen}, \&
  {Yasukawa}}]{Hansen1974}
{Hansen}, R.~T., {Garcia}, C.~J., {Hansen}, S.~F., \& {Yasukawa}, E. 1974,
  \pasp, 86, 500

\bibitem[{{Harra} {et~al.}(2007){Harra}, {Hara}, {Imada}, {Young}, {Williams},
  {Sterling}, {Korendyke}, \& {Attrill}}]{Harra2007}
{Harra}, L.~K., {Hara}, H., {Imada}, S., {et~al.} 2007, \pasj, 59, S801

\bibitem[{{Harra} \& {Sterling}(2001)}]{Harra2001}
{Harra}, L.~K., \& {Sterling}, A.~C. 2001, \apjl, 561, L215

\bibitem[{{Harrison} \& {Lyons}(2000)}]{Harrison2000}
{Harrison}, R.~A., \& {Lyons}, M. 2000, \aap, 358, 1097

\bibitem[{{Howard} \& {Vourlidas}(2018)}]{Howard2018}
{Howard}, R.~A., \& {Vourlidas}, A. 2018, \solphys, 293, 55

\bibitem[{{Howard} {et~al.}(2008){Howard}, {Moses}, {Vourlidas}, {Newmark},
  {Socker}, {Plunkett}, {Korendyke}, {Cook}, {Hurley}, {Davila}, {Thompson},
  {St Cyr}, {Mentzell}, {Mehalick}, {Lemen}, {Wuelser}, {Duncan}, {Tarbell},
  {Wolfson}, {Moore}, {Harrison}, {Waltham}, {Lang}, {Davis}, {Eyles},
  {Mapson-Menard}, {Simnett}, {Halain}, {Defise}, {Mazy}, {Rochus}, {Mercier},
  {Ravet}, {Delmotte}, {Auchere}, {Delaboudiniere}, {Bothmer}, {Deutsch},
  {Wang}, {Rich}, {Cooper}, {Stephens}, {Maahs}, {Baugh}, {McMullin}, \&
  {Carter}}]{Howard2008}
{Howard}, R.~A., {Moses}, J.~D., {Vourlidas}, A., {et~al.} 2008, \ssr, 136, 67

\bibitem[{{Hudson} {et~al.}(1996){Hudson}, {Acton}, \& {Freeland}}]{Hudson1996}
{Hudson}, H.~S., {Acton}, L.~W., \& {Freeland}, S.~L. 1996, \apj, 470, 629

\bibitem[{{Imada} {et~al.}(2007){Imada}, {Hara}, {Watanabe}, {Kamio}, {Asai},
  {Matsuzaki}, {Harra}, \& {Mariska}}]{Imada2007}
{Imada}, S., {Hara}, H., {Watanabe}, T., {et~al.} 2007, \pasj, 59, S793

\bibitem[{{Janvier} {et~al.}(2015){Janvier}, {Aulanier}, \&
  {D{\'e}moulin}}]{Janvier2015}
{Janvier}, M., {Aulanier}, G., \& {D{\'e}moulin}, P. 2015, \solphys, 290, 3425

\bibitem[{{Jin} {et~al.}(2009){Jin}, {Ding}, {Chen}, {Fang}, \&
  {Imada}}]{Jin2009}
{Jin}, M., {Ding}, M.~D., {Chen}, P.~F., {Fang}, C., \& {Imada}, S. 2009, \apj,
  702, 27

\bibitem[{{Kamio} {et~al.}(2010){Kamio}, {Hara}, {Watanabe}, {Fredvik}, \&
  {Hansteen}}]{Kamio2010}
{Kamio}, S., {Hara}, H., {Watanabe}, T., {Fredvik}, T., \& {Hansteen}, V.~H.
  2010, \solphys, 266, 209

\bibitem[{{Kosugi} {et~al.}(2007){Kosugi}, {Matsuzaki}, {Sakao}, {Shimizu},
  {Sone}, {Tachikawa}, {Hashimoto}, {Minesugi}, {Ohnishi}, {Yamada}, {Tsuneta},
  {Hara}, {Ichimoto}, {Suematsu}, {Shimojo}, {Watanabe}, {Shimada}, {Davis},
  {Hill}, {Owens}, {Title}, {Culhane}, {Harra}, {Doschek}, \&
  {Golub}}]{Kosugi2007}
{Kosugi}, T., {Matsuzaki}, K., {Sakao}, T., {et~al.} 2007, \solphys, 243, 3

\bibitem[{{Lemen} {et~al.}(2012){Lemen}, {Title}, {Akin}, {Boerner}, {Chou},
  {Drake}, {Duncan}, {Edwards}, {Friedlaender}, {Heyman}, {Hurlburt}, {Katz},
  {Kushner}, {Levay}, {Lindgren}, {Mathur}, {McFeaters}, {Mitchell}, {Rehse},
  {Schrijver}, {Springer}, {Stern}, {Tarbell}, {Wuelser}, {Wolfson}, {Yanari},
  {Bookbinder}, {Cheimets}, {Caldwell}, {Deluca}, {Gates}, {Golub}, {Park},
  {Podgorski}, {Bush}, {Scherrer}, {Gummin}, {Smith}, {Auker}, {Jerram},
  {Pool}, {Soufli}, {Windt}, {Beardsley}, {Clapp}, {Lang}, \&
  {Waltham}}]{Lemen2012}
{Lemen}, J.~R., {Title}, A.~M., {Akin}, D.~J., {et~al.} 2012, \solphys, 275, 17

\bibitem[{{Mandrini} {et~al.}(2007){Mandrini}, {Nakwacki}, {Attrill}, {van
  Driel-Gesztelyi}, {D{\'e}moulin}, {Dasso}, \& {Elliott}}]{Mandrini2007}
{Mandrini}, C.~H., {Nakwacki}, M.~S., {Attrill}, G., {et~al.} 2007, \solphys,
  244, 25

\bibitem[{{Miklenic} {et~al.}(2011){Miklenic}, {Veronig}, {Temmer},
  {M{\"o}stl}, \& {Biernat}}]{Miklenic2011}
{Miklenic}, C., {Veronig}, A.~M., {Temmer}, M., {M{\"o}stl}, C., \& {Biernat},
  H.~K. 2011, \solphys, 273, 125

\bibitem[{{Pesnell} {et~al.}(2012){Pesnell}, {Thompson}, \&
  {Chamberlin}}]{Pesnell2012}
{Pesnell}, W.~D., {Thompson}, B.~J., \& {Chamberlin}, P.~C. 2012, \solphys,
  275, 3

\bibitem[{{Robbrecht} \& {Wang}(2010)}]{Robbrecht2010}
{Robbrecht}, E., \& {Wang}, Y.-M. 2010, \apjl, 720, L88

\bibitem[{{Scherrer} {et~al.}(2012){Scherrer}, {Schou}, {Bush}, {Kosovichev},
  {Bogart}, {Hoeksema}, {Liu}, {Duvall}, {Zhao}, {Title}, {Schrijver},
  {Tarbell}, \& {Tomczyk}}]{Scherrer2012}
{Scherrer}, P.~H., {Schou}, J., {Bush}, R.~I., {et~al.} 2012, \solphys, 275,
  207

\bibitem[{{Sterling} \& {Hudson}(1997)}]{Sterling1997}
{Sterling}, A.~C., \& {Hudson}, H.~S. 1997, \apjl, 491, L55

\bibitem[{{Tappin}(2006)}]{Tappin2006}
{Tappin}, S.~J. 2006, \solphys, 233, 233

\bibitem[{{Temmer} {et~al.}(2017){Temmer}, {Thalmann}, {Dissauer}, {Veronig},
  {Tschernitz}, {Hinterreiter}, \& {Rodriguez}}]{Temmer2017}
{Temmer}, M., {Thalmann}, J.~K., {Dissauer}, K., {et~al.} 2017, \solphys, 292,
  93

\bibitem[{{Thompson} {et~al.}(2000){Thompson}, {Cliver}, {Nitta},
  {Delann{\'e}e}, \& {Delaboudini{\`e}re}}]{Thompson2000}
{Thompson}, B.~J., {Cliver}, E.~W., {Nitta}, N., {Delann{\'e}e}, C., \&
  {Delaboudini{\`e}re}, J.-P. 2000, \grl, 27, 1431

\bibitem[{{Thompson} {et~al.}(1998){Thompson}, {Plunkett}, {Gurman}, {Newmark},
  {St.~Cyr}, \& {Michels}}]{Thompson1998}
{Thompson}, B.~J., {Plunkett}, S.~P., {Gurman}, J.~B., {et~al.} 1998, \grl, 25,
  2465

\bibitem[{{Tian} {et~al.}(2012){Tian}, {McIntosh}, {Xia}, {He}, \&
  {Wang}}]{Tian2012}
{Tian}, H., {McIntosh}, S.~W., {Xia}, L., {He}, J., \& {Wang}, X. 2012, \apj,
  748, 106

\bibitem[{{Vanninathan} {et~al.}(2018){Vanninathan}, {Veronig}, {Dissauer}, \&
  {Temmer}}]{Vanninathan2018}
{Vanninathan}, K., {Veronig}, A.~M., {Dissauer}, K., \& {Temmer}, M. 2018,
  \apj, 857, 62

\bibitem[{{Vourlidas} {et~al.}(2010){Vourlidas}, {Howard}, {Esfandiari},
  {Patsourakos}, {Yashiro}, \& {Michalek}}]{Vourlidas2010}
{Vourlidas}, A., {Howard}, R.~A., {Esfandiari}, E., {et~al.} 2010, \apj, 722,
  1522

\bibitem[{{Vourlidas} {et~al.}(2000){Vourlidas}, {Subramanian}, {Dere}, \&
  {Howard}}]{Vourlidas2000}
{Vourlidas}, A., {Subramanian}, P., {Dere}, K.~P., \& {Howard}, R.~A. 2000,
  \apj, 534, 456

\bibitem[{{Webb} {et~al.}(2000){Webb}, {Cliver}, {Crooker}, {Cry}, \&
  {Thompson}}]{Webb2000}
{Webb}, D.~F., {Cliver}, E.~W., {Crooker}, N.~U., {Cry}, O.~C.~S., \&
  {Thompson}, B.~J. 2000, \jgr, 105, 7491

\bibitem[{{Webb} {et~al.}(1996){Webb}, {Howard}, \& {Jackson}}]{Webb1996}
{Webb}, D.~F., {Howard}, R.~A., \& {Jackson}, B.~V. 1996, in American Institute
  of Physics Conference Series, Vol. 382, American Institute of Physics
  Conference Series, ed. D.~{Winterhalter}, J.~T. {Gosling}, S.~R. {Habbal},
  W.~S. {Kurth}, \& M.~{Neugebauer}, 540--543

\bibitem[{{Yashiro} {et~al.}(2004){Yashiro}, {Gopalswamy}, {Michalek},
  {St.~Cyr}, {Plunkett}, {Rich}, \& {Howard}}]{Yashiro2004}
{Yashiro}, S., {Gopalswamy}, N., {Michalek}, G., {et~al.} 2004, Journal of
  Geophysical Research (Space Physics), 109, A07105

\bibitem[{{Young} {et~al.}(2007){Young}, {Del Zanna}, {Mason}, {Dere}, {Landi},
  {Landini}, {Doschek}, {Brown}, {Culhane}, {Harra}, {Watanabe}, \&
  {Hara}}]{Young2007}
{Young}, P.~R., {Del Zanna}, G., {Mason}, H.~E., {et~al.} 2007, \pasj, 59, S857

\bibitem[{{Zarro} {et~al.}(1999){Zarro}, {Sterling}, {Thompson}, {Hudson}, \&
  {Nitta}}]{Zarro1999}
{Zarro}, D.~M., {Sterling}, A.~C., {Thompson}, B.~J., {Hudson}, H.~S., \&
  {Nitta}, N. 1999, \apjl, 520, L139

\bibitem[{{Zhukov} \& {Auch{\`e}re}(2004)}]{Zhukov2004}
{Zhukov}, A.~N., \& {Auch{\`e}re}, F. 2004, \aap, 427, 705

\end{thebibliography}

\end{document}